# Estimating the effects of sea level rise on coupled estuarine nitrogen cycling processes through comparative network analysis

David E. Hines[1, 2,*], Jessica A. Lisa[3], Bongkeun Song[3], Craig R. Tobias[4], Stuart R. Borrett[1, 2]

[1] Department of Biology & Marine Biology, University of North Carolina Wilmington, Wilmington, North Carolina

[2] Center for Marine Science, University of North Carolina Wilmington, Wilmington, North Carolina

[3] Department of Biological Sciences, Virginia Institute of Marine Science, Gloucester Point, Virginia

[4] Department of Marine Sciences, University of Connecticut, Gorton, Connecticut

Corresponding address: deh9951@uncw.edu
Corresponding telephone: 910-962-2411
Corresponding fax: 910-962-4066

**Abstract**

Nitrogen (N) removal from estuaries is driven in part by sedimentary microbial processes. The processes of denitrification and anaerobic ammonium oxidation (anammox) remove N from estuaries by producing di-nitrogen gas, and each can be coupled to N recycling pathways such as nitrification and dissimilatory nitrate reduction to ammonium (DNRA). Environmental conditions in estuaries influence sedimentary N cycling processes; therefore, seawater intrusion may affect the coupling of N cycling processes in the freshwater portions of estuaries. This study investigated the potential effects of seawater intrusion on these process couplings through a comparative modeling approach. We applied environ analysis, a form of ecological network analysis, to two N cycling mass-balance network models constructed at freshwater (oligohaline) and saltwater (polyhaline) sites in the Cape Fear River Estuary, North Carolina. We used a space-for-time substitution to predict the effects of seawater intrusion on the sedimentary N cycle. Further, we conducted an uncertainty analysis using linear inverse modeling to evaluate the effects of parameterization uncertainty on model results. Nitrification coupled to both denitrification and anammox was 2.5 times greater in the oligohaline model, while DNRA coupled to anammox was 2.7 times greater in the polyhaline model. However, the total amount of $N_2$ gas produced relative to the nitrogen inputs to each network was 4.7% and 4.6% at the oligohaline and polyhaline sites, respectively. These findings suggest that changes in water chemistry from seawater intrusion may favor direct over coupled nitrogen removal, but may not substantially change the N removal capacity of the sedimentary microbial processes.

# 1 Introduction

Estuarine sediments support microbial communities that provide important ecosystem services ranging from the decomposition of organic material to the recycling and removal of nutrients (Costanza et al. 1998; Barbier et al. 2011). The microbial communities that provide these ecosystem services can vary with environmental conditions such as temperature, nutrient availability, and salinity (Bouvier and del Giorgio 2002; Bernhard et al. 2007, Baron et al. 2013). However, the environmental conditions in estuarine ecosystems may undergo substantial changes in the future resulting from continued urban development and global climate change. Specifically, the dredging of estuarine channels and the addition of canals to accommodate shipping traffic may lead to seawater intrusion into the freshwater portions of estuaries (Newport 1977; Hackney and Yelverton 1990; Zhang et al. 2012). Further, predicted sea level rise over the next century (IPCC 2007) may intensify the effects of the shifts in water chemistry associated with seawater intrusion. For example, increased salinity can facilitate the release of ammonium ($NH_4^+$) from estuarine sediments to the water column where it is available to planktonic algae (Gardner et al. 1991; Hou et al. 2003), and can also repress the removal rates of nitrogen (N) by microbial organisms (Dong et al. 2000). These environmental shifts may substantially hinder the ecosystem services that estuaries provide, including N removal (Craft et al. 2008). The objectives of this work were to describe, quantify, and predict the potential effects of seawater intrusion on some of these ecosystem services. Specifically, we investigated how seawater intrusion might alter the interactions between microbially mediated processes in the estuarine N cycle.

The removal of nutrients from estuarine sediments through microbially mediated processes can have important implications for the health of estuaries (Pickney et al. 2001). Because primary production in estuaries is typically limited by the availability of N (Ryther and Dunstan 1971; Howarth and Marino 2006), N removal processes can help to alleviate the effects of eutrophication caused by anthropogenic nutrient loading (Anderson et al. 2002). Two microbial processes in sediments convert biologically available forms of N to biologically unavailable di-nitrogen ($N_2$) gas in estuaries: denitrification and anaerobic ammonium oxidation (anammox). Denitrification combines two molecules of nitrate ($NO_3^-$) to produce $N_2$, while anammox combines one molecule of $NH_4^-$ with one molecule of nitrite ($NO_2^-$) to convert N to $N_2$. In addition, microbial processes can transform N between different species. For example, dissimilatory nitrate reduction to ammonium (DNRA) transforms $NO_3^-$ through $NO_2^-$ to produce $NH_4^+$, while nitrification transforms $NH_4^+$ through $NO_2^-$ to produce $NO_3^-$ (Francis et al. 2007). These N transformations can play a key role in regulating the availability of N for primary producers, as well as for the removal processes of denitrification and anammox.

Denitrification and anammox can be either uncoupled (direct) or coupled to N transformation processes. Direct N removal processes produce $N_2$ from N in the form it enters the estuary, while coupled removal processes consume the products of microbial N transformation processes (Jenkins and Kemp 1984). For example, denitrification and anammox can utilize $NO_3^-$ and $NO_2^-$ produced by nitrification, respectively. Anammox can also utilize $NH_4^+$ and $NO_2^-$ produced by DNRA. The strength of coupling between N removal and transformation processes relative to direct removal processes can have important implications for N residence time and the N removal capacity of estuaries (Thamdrump and Dalsgaard 2002; Seitzinger et al. 2006; Santoro 2010).

Alterations in the coupling strength of N removal and transformation processes may exacerbate eutrophication. Seawater intrusion induced release of $NH_4^+$ from estuarine sediments to the water column can provide nutrients for N limited algae. Further, because denitrification is the dominant N removal pathway, a strong coupling between nitrification and denitrification is necessary to remove the $NH_4^+$ before it can be assimilated into algal biomass. However, environmental conditions affect the rates of microbial N transformation processes. For example, nitrification rates typically decrease at elevated salinities due to sulfide inhibition (Joye and Hollibaugh 1995; Rysgaard et al. 1999), while DNRA activity tends to increase along a salinity gradient (Giblin et al. 2010). Under elevated salinity conditions, estuaries could experience weakened coupling of microbial N processes and may be at higher risk for prolonged eutrophication. In this study we used network modeling and analysis to estimate the potential effects of seawater intrusion on the coupling of microbial N cycling processes and evaluate the possible implications of these effects on estuarine N removal.

Ecological Network Analysis (ENA) provides a means to evaluate the strength of coupling between N removal and transformation processes, along with the potential effects of environmental changes on these relationships (Hines et al. 2012). Ecological networks provide a whole-ecosystem perspective in which thermodynamically conserved material and material fluxes comprise network nodes and edges, respectively. ENA is a set of analyses derived from economic input-output analyses that are applied to mass-balanced network models to evaluate the flow of energy-matter through an ecosystem (Hannon 1973; Patten et al. 1976; Fath and Patten 1999; Ulanowicz 2004). Environ analysis, a form of flow analysis in ENA, is used to partition the flow of energy-matter in an ecosystem network to track material moving through the ecosystem (Patten 1978; 1982; Fath and Patten 1999). Comparisons among different parameterizations of ecological networks using ENA tools can provide insight into how differences in network organization can affect the system behavior. For example, Christian et al. (2005) compare ecological trophic networks of different estuaries at different seasons to draw conclusions about the effects of stress on the system, while Heymans et al. (2002) compare ecological networks of two Florida marshes to determine relative system maturity. We applied environ analysis in ENA to estimate the coupling of N transformation and removal processes in N cycling networks (Hines et al. 2012).

To determine the potential equilibrium effect of seawater intrusion on the sedimentary N cycle, we compared two N cycling networks parameterized at freshwater (oligohline) and saltwater (polyhaline) sites, respectively, in the Cape Fear River Estuary (CFRE), North Carolina. We used environ analysis to evaluate the strengths of the coupling of nitrification to denitrification as well as of nitrification and DNRA to anammox at each site. Because of the negative effects of seawater on the process of nitrification, we hypothesized that nitrification coupled to denitrification and nitrification coupled to anammox would be lower in the polyhaline network. However, we hypothesized that DNRA coupled to anammox would be higher in the polyhaline network due to the resilience of DNRA to seawater conditions. To evaluate the effect of seawater intrusion on the microbial N cycle, we used a space-for-time substitution by comparing the model results between the oligohaline and polyhaline sites (Pickett 1989). This approach assumes that, near equilibrium, the oligohaline site will resemble the polyhaline site after the seawater has replaced the freshwater. Further, we evaluated the robustness of the model results to uncertainty in network parameterization by conducting an uncertainty analysis. Specifically, we used linear inverse modeling and a Monte Carlo approach to construct a range of plausible models and compared the distributions of model calculations at each site. This study 1) synthesized disparate measurements to estimate the direct and coupled N removal from denitrification and anammox, 2) tested hypotheses about the effect of seawater intrusion on process coupling, and 3) evaluated the N removal capacity of estuarine sediments under different salinity regimes.

## 2 Methods

*2.1 Network construction*

In this work, we compared two N cycling networks for the CFRE, one at an oligohaline site (Oligo) and one at a polyhaline site (Poly; Fig. 1). The Oligo network was constructed by Hines et al. (2012) and models a section of river called Horseshoe Bend (34° 14' 37.464'' N, 77° 58' 11.280'' W), which typically experiences salinities ranging from 0.1 to 5.0, with occasional salinities as high as 8 (Mallin et al. 2009, 2010). We constructed a second N cycling network for comparison at the Poly site, a section of river at channel marker 35 (34° 2' 2.688'' N, 77° 56' 21.948'' W) with a mean salinity consistently above 10 (Mallin et al. 2009, 2010). Both networks had identical topologies to facilitate comparison (e.g., Baird et al. 1991), but the parameter values representing the magnitudes of N storages and fluxes for each network varied according to differences in the N cycle observed at each site. The mean of direct measurements of N storages and transformations at each site were used to parameterize the networks whenever possible; values of fluxes reported in the literature for similar sites were used when direct measurements were not available (Table 1). The models were parameterized to represent the average conditions at the site during a single day in the summer months (June – August, 2008-2009).

Each network represents a control volume consisting of adjacent 1 cm$^3$ segments of the water column and sediment at the water-sediment interface at the Oligo and Poly sites, respectively. As a prerequisite to conducting ENA, each network should encompass all aspects of the ecosystem of interest through representation as storages or fluxes (Fath et al. 2007). The small scale of these models enabled the networks to focus on the microbial processes involved in N cycling by considering the N contributions of macroorganisms and large detritus as boundary inputs and outputs to the modeled volume (Hines et al. 2012). Specifically, the contributions of these ecosystem components to the N cycle in the networks were through dissolved N inputs to the control volume, which were accounted for in direct measurements of N inputs to each site (Table 1). These models provide a small-scale perspective of the interactions between microbial processes, and thus are useful for observing changes in these interactions.

Each control volume was assumed to be at a steady state over the duration of one summer day. This assumption is necessary to conduct the mathematics of ENA and has been applied to network models of estuarine ecosystems in the past (Baird and Heymans 1996; Christian and Thomas 2003).

In each control volume, N was divided into pools of ammonium ($NH_4$), nitrate and nitrite ($NO_X$), the N stored in microbial biomass (M), and a combination of dissolved and particulate organic N (ON). Each N pool was assigned a node in the network and pools were repeated in the water column (W-) and sediment (S-), yielding a total of eight network nodes (Fig. 2). N transformation processes were measured in units of nmol N cm$^{-3}$ d$^{-1}$ and were used to guide the construction of network links; a detailed description of the storages and internal fluxes in the networks as well as a complete justification for each element of the network design can be found in Hines et al. (2012). The values used for each network flux and storage, and the sources from which these values were obtained, can be found in Table 1.

*2.2 Model evaluation*

The quality of the information used to develop a model is directly related to the accuracy and validity of the model results. As an initial tool to assess model quality, Hines et al. (2012)

classified the quality of information in the Oligo model using an information-ranking rubric developed by Costanza (1992). According to this rubric, high quality information comes from direct measurement, medium quality data comes from calculations based on direct measurement, and low quality data comes from plausible estimation. We applied the Costanza (1992) rubric in a similar manner to qualify the quality of information used in the construction of the Poly model (Table 2).

*2.3 Matrix representation*

The networks constructed at each site (Fig. 2) were isomorphically represented as two sets of matrices and vectors to facilitate the calculations involved in ENA. The internal N fluxes between nodes for each network were represented in the flow matrix $\mathbf{F} = (f_{ij})$ where $i$ refers to matrix rows (sink compartment) and $j$ refers to matrix columns (source compartment). System inputs and losses to each network were captured in the input vector $z = (z_i)$ and output vector, $y = (y_j)$, respectively. The N stored in each pool was represented by the storage vector $x = (x_i)$. Flow of N in these matrices and vectors was oriented from columns ($j$) to rows ($i$). Complete models formatted according to the Scientific Committee on Ocean Research standards (SCOR files; Ulanowicz and Kay 1991) can be found in both print and electronic form in the Web Appendix. A detailed description of the matrix and vector representation used here is provided in Hines et al. (2012).

*2.4 Ecological network analysis (ENA)*

ENA is applied to steady state ecosystem networks to characterize the flow of energy-matter through the system (Patten et al. 1976; Ulanowicz 1986). ENA consists of several different mathematical analyses including flow and environ analyses; detailed reviews of ENA can be found in Fath and Patten (1999), Ulanowicz (2004), and Schramski et al. (2011). This study applied ENA to the Oligo and Poly N networks using the *enaR* package for R (Lau et al. 2012). Following targeted mass-balancing during the model construction, the difference between the input and output fluxes of each network node were less the 5%. Thus, the networks were considered to be at steady-state. The results of ENA were used to estimate the coupling of microbial processes at the Oligo and Poly sites, and were compared to make predictions about the potential effects of sea level rise on the microbial nitrogen cycle in estuarine sediments.

*2.4.1 Flow analysis*

Flow analysis is used to determine how much material travels across the different pathways and through the different nodes in an ecosystem network (Finn 1980; Hannon 1985). Analyses exist for both the time-backward (input) and time-forward (output) flow orientations that provide different perspectives of the same system (Schramski et al. 2011; Borrett and Freeze 2011). The throughflow of a given node in a network $T$ is calculated as $T_i^{in} = \sum_{j=1}^{n} f_{ij} + z_i$ for the input orientation or $T_j^{out} = \sum_{i=1}^{n} f_{ij} + y_j$ for the output orientation where *n* is the number of nodes in the network. At steady state, $T_i^{in} = T_j^{out} = T$ for each network node (Schramski et al. 2011). The work presented here utilizes the input-oriented analysis, focusing on where N enters and travels through the system before it exits the control volume for each network.

In flow analysis, several network-level indicators are used to characterize the movement of energy-matter through an ecosystem. One commonly used indicator, total system throughflow (TST), is the sum of all activity across all nodes and is calculated as $TST = \sum_{j=1}^{n} T_j$. Another commonly used network-level indicator, the Finn Cycling Index (FCI), is used to determine how much material in a specific network node leaves that node and returns to it at least once before exiting the network (Finn 1980). FCI may underestimate cycling within systems by approximately 14.2% because it omits compound pathways, but is far less computationally intensive and more commonly used than the comprehensive cycling index (CCI) proposed by Allesina and Ulanowicz (2004), which accounts for compound pathways. We calculated TST and the FCI to quantify and characterize the total energy-matter flux within the Oligo and Poly networks at a broad scale, and inflated the FCI calculations by 14.2% as suggested by Allesina and Ulanowicz (2004) to estimate CCI values. ENA also allows for finer-scale analysis of ecosystem activity using a pathway decomposition of throughflow.

The direct flow intensity matrix **G** provides the magnitudes of fluxes traveling from one node to another over a path length of one edge. In the input-oriented direction (denoted by ′), the direct flow intensity matrix is defined as $\mathbf{G}' = (g'_{ij}) = f_{ij}/T_i$ where $T_i$ is the throughflow in the receiving node. The input-oriented integral flow intensity matrix $\mathbf{N}' = (n'_{ij})$ represents the flow from one network node to another across all path lengths. $\mathbf{N}'$ is the sum of the boundary ($\mathbf{G}'^0$), direct ($\mathbf{G}'^1$), and indirect ($\mathbf{G}'^m$) flow intensity matrices such that

$$\mathbf{N}' = \underbrace{\mathbf{G}'^0}_{\text{boundary}} + \underbrace{\mathbf{G}'^1}_{\text{direct}} + \underbrace{\mathbf{G}'^2 + \cdots + \mathbf{G}'^m + \cdots}_{\text{indirect}}$$

where $m$ is the path length. The identity $\mathbf{N}' = (\mathbf{G}'^0 - \mathbf{G}'^1)^{-1}$ is used to calculate an exact solution for the integral flow intensity matrix because $\mathbf{N}'$ is a convergent series in steady state networks (Ulanowicz 2004; Borrett et al. 2010). Network ecologists have used flow analysis to identify emergent properties in ecosystems such as the dominance of indirect effects and network homogenization, which are hypothesized to be general properties of all ecosystem networks (Patten et al. 1990; Fath and Patten 1999; Salas and Borrett 2011). A detailed description of flow analysis can be found in the literature (Patten et al. 1976; Fath and Patten1999; Borrett et al. 2010; Schramski et al. 2011).

*2.4.2 Environ analysis*

Environ analysis partitions the quantified flows in a network to show where material comes from after it exits the network (time-backward, input orientation) or where material goes after it enters the network (time-forward, output orientation; Patten 1978; 1981; 1982). The environs produced by environ analysis are non-overlapping subnetworks that can be summed to recover the original network (Patten, 1978). Formally, input oriented environs $e'_{ijk}$ are calculated by placing elements of the integral flow intensity matrix along the principal diagonal of a square matrix $\delta'_{\ell j}$, then multiplying elements of the direct flow intensity matrix $g'_{i\ell}$ by the resulting matrix such that $e'_{ijk} = g'_{i\ell} \times \delta'_{\ell j}$ where

$$\delta'_{\ell j} = \begin{cases} n'_{ik} & \text{if } \ell = j \\ 0 & \text{if } \ell \neq j \end{cases}$$

The environ analysis applied in this work utilizes realized input environs $\bar{e}'_{ijk}$, which are unit environs rescaled by the observed system boundary flows (Whipple et al. 2007; Borrett and Freeze, 2011). Realized input environs are calculated by multiplying the input oriented environs by the corresponding network output $y_k$ so that $\bar{e}'_{ijk} = y_k \times e'_{ijk}$. Whipple et al. (2007) apply comparative realized environ analysis to a series of seasonal estuarine N cycling models to examine changes in compartmental contribution to flow over time. More recently, realized environ analysis has been adapted to examine the internal network activity involved with individual boundary fluxes. Specifically, Hines et al. (2012) use realized environ analysis of a N network to estimate the coupling of microbial N transformation processes to the N removal processes of denitrification and anammox in estuarine sediments.

*2.4.3 Coupling quantification*

The realized input environs generated by environ analysis were used to estimate the coupling of nitrification to denitrification as well as nitrification to anammox and DNRA to anammox. Fig. 3 shows an example of how the denitrification environs for each network were used to calculate nitrification coupled to denitrification. N in the S-NO$_X$ node was assumed to have a probability of exiting the network through the denitrification pathway ($p_d$) equal to the proportion of N involved in the denitrification pathway relative to all N exiting the node in the realized denitrification environ so that

$$p_d = \frac{A}{A+B+C+D}$$

where $A$ is the magnitude of the denitrification flux, $B$ is the uptake of NO$_X$ by microbes in the sediments, $C$ is the movement of NO$_X$ from the sediments to the water column, and $D$ is the conversion of $NO_3^-$ and $NO_2^-$ to $NH_4^+$ in the sediments through DNRA (Fig. 3). The amount of N involved in nitrification coupled to denitrification ($Coupled_{nd}$) was calculated by multiplying the amount of N crossing the nitrification pathway in the denitrification environ ($X$) by the probability of N in the S-NO$_X$ node exiting the network through denitrification ($p_d$) so that

$$Coupled_{nd} = X \times p_d = \frac{X \times A}{A+B+C+D}$$

The strength of the coupling of nitrification to denitrification ($CS_{nd}$) was obtained by dividing the coupled nitrification to denitrification by the total denitrification removal, resulting in

$$CS_{nd} = \frac{X \times A}{A+B+C+D} \times \frac{1}{A} = \frac{X}{A+B+C+D}$$

The strength of the coupling ($CS_{nd}$) was then multiplied by one hundred to determine the percentage of nitrification coupled to denitrification. Similar calculations were used for the S-NH$_4$ and S-NO$_X$ nodes in the realized anammox environs to determine nitrification coupled to anammox and DNRA coupled to anammox.

*2.4.4 Nitrogen removal efficiency*

The magnitudes of microbial N removal processes with respect to the N inputs at each site revealed the relative ability of the microbial communities to utilize the available resources. Coupled and direct N removal processes at the Oligo and Poly sites were scaled to the inputs of each network by dividing each removal process ($R_p$) by the sum of the appropriate input vector (z) such that the relative process magnitudes were $\frac{R_p}{\sum_{i=1}^{n} z}$.

*2.5 Uncertainty analysis*

The model quantifications of coupling and N removal efficiency at the Oligo and Poly sites are based on calculations that rely on the network parameterization at each location. However, the data used to assign flow magnitudes to individual fluxes were averaged across multiple summer seasons and, therefore, contained uncertainty that differed among each flux. Further, some model parameters were obtained from literature measurements in similar estuaries or by mass balance, adding to the uncertainty in the parameterization of each network. These uncertainties, which are common in models (Oreskes et al. 1994), imply that a range of plausible networks and associated coupling quantifications exists for each of the Oligo and Poly sites (e.g., Borrett and Osidele 2007).

To evaluate the robustness of our model conclusions to these parameter uncertainties, we performed an uncertainty analyses. We used a linear inverse modeling approach based on the techniques presented by Kones et al. (2009) to create 10,000 plausible model parameterizations for each site. We used the *limSolve* package for R (Soetaert 2009) to execute the analysis.

Plausible models were considered to 1) be at steady state and 2) contain parameters with values within the range of certainty for each network flux.

We used a stratified sampling technique for the uncertainty analysis. High quality parameters under the Constanza (1992) rubric, which consisted of direct measurements from each site, were restricted to within one standard deviation of the mean measured value. Medium quality parameters were restricted to within a percentage of the value in the original networks (Fig. 2) equal to the largest percent variation observed in the high quality data (±47%), while low quality parameters were allowed to vary by either ±50% of the original network values for a first analysis or ±100% for a second analysis. We compared the results of the two levels of uncertainty in the low quality parameter estimates to determine the relative impact of this uncertainty. The classification of each network flux along with the percentage of disturbance used for plausible model construction can be found in the Table 2.

For each model realization in the uncertainty analysis, we performed the ENA and coupling analysis. This let us determine the 95% confidence intervals for the couplings and removal capacities at the Olio and Poly sites and generally estimate the robustness of the model results to the underlying model uncertainty.

# 3 Results
## 3.1 Model evaluation
The Oligo model was constructed from 26% high, 51% medium, and 23% low quality information according to the Costanza (1992) evaluation rubric (Hines et al. 2012). This distribution implies that 77% of the information used in the model construction is based on empirical measurements. The Poly model displayed the same 26%, 51%, and 23% high, medium, and low quality distribution among the ranking categories as the Oligo model. Further, the quality of parameters used for each network flux was identical between the two models, facilitating their comparison (Table 2).

## 3.2 Flow analysis
The flow of N through the networks differed in both magnitude and organization between the Oligo and Poly sites. The TST at the Oligo site (7088.7 nmol N $cm^{-3}d^{-1}$) was 33.1% higher than TST at the Poly site (5326.8 nmol N $cm^{-3}d^{-1}$). Similarly, the recycling at the Oligo site (FCI = 0.20; estimated CCI = 0.23) was 17.7% higher than the Poly site (FCI = 0.17; estimated CCI = 0.19).

## 3.3 Environ analysis
Environ analysis generated twelve realized input environs for each network, one for every network boundary output. Each environ revealed the amount of N traveling across input and internal pathways that was associated with a specific output boundary flux for a given network. In the denitrification environs, 76.9 nmol N $cm^{-3}d^{-1}$ was involved in sediment nitrification (S-$NH_4$ $\rightarrow$ S-$NO_X$) at the Oligo site, while only 25.1 nmol N $cm^{-3}d^{-1}$ was involved in sediment nitrification at the Poly site (Fig. 4). In the anammox environs, 1.2 nmol N $cm^{-3}d^{-1}$ and 0.1 nmol N $cm^{-3}d^{-1}$ were involved in sediment nitrification and DNRA (S-$NO_X$ $\rightarrow$ S-$NH_4$), respectively, at the Oligo site; 0.4 nmol N $cm^{-3}d^{-1}$ and 0.2 nmol N $cm^{-3}d^{-1}$ were involved in the same processes, respectively, at the Poly site (Fig. 5).

## 3.4 Calculation of coupling
Environ results were used to calculate the coupling of microbial N processes at the Oligo and Poly sites. At the Oligo site, an estimated 43.5% (74.8 nmol N $cm^{-3}$ $d^{-1}$) of denitrification was coupled to nitrification, while the remaining 56.5% (97.2 nmol N $cm^{-3}$ $d^{-1}$) was a result of direct denitrification. The strength of the coupling between nitrification and denitrification decreased between the Oligo and Poly sites. Coupled nitrification to denitrification was responsible for just 18.0% (24.6 nmol N cm$^{-3}$ d$^{-1}$) of denitrification activity at the Poly site, while direct denitrification was responsible for 82.0% (112.1 nmol N cm$^{-3}$ d$^{-1}$) of denitrification activity (Fig. 6).

Direct anammox was greater than coupled anammox at both study sites; however, differences in the strength of coupling between the Oligo and Poly sites were observed. Nitrification coupled to anammox and DNRA coupled to anammox at the Oligo site were responsible for 22.7% (1.1 nmol N cm$^{-3}$ d$^{-1}$) and 1.8% (0.1 nmol N cm$^{-3}$ d$^{-1}$) of anammox activity, respectively. The remaining 75.4% (3.8 nmol N cm$^{-3}$ d$^{-1}$) of anammox activity was a result of direct anammox. At the Poly site, however, the strength of nitrification coupled to anammox weakened to 9.6% (0.3 nmol N cm$^{-3}$ d$^{-1}$), while the strength of DNRA coupled to anammox increased to 4.8% (0.2 nmol N cm$^{-3}$ d$^{-1}$). The remaining 85.6% (3.1 nmol N cm$^{-3}$ d$^{-1}$) of anammox activity was a result of direct anammox (Fig. 6).

The process coupling values for the Oligo network presented in this study were not identical to values published for the same network in previous work (Hines et al. 2012) as a result of differences in the software used to conduct the analyses. However, all of the differences were less than 2% of the flux magnitudes and do not affect the conclusions of this study.

*3.5 Nitrogen removal efficiency*

Although differences were observed in the strength of coupling of microbial processes between the Oligo and Poly sites, little change was seen in the ability of the microbial communities to remove N relative to the N inputs at each site. The amount of N removed by denitrification and anammox was higher at the Oligo site (177.0 nmol N cm$^{-3}$ d$^{-1}$) than the Poly site (140.2 nmol N cm$^{-3}$ d$^{-1}$), but relative to the total N inputs into the Oligo and Poly networks, total $N_2$ production from these two processes was 4.7% and 4.6%, respectively (Fig. 7)

*3.6 Uncertainty analysis*

A Monte Carlo approach was used in conjunction with linear inverse modeling to generate 10,000 plausible networks at each site that resulted in calculated distributions of process coupling and N removal capacity. The mean and standard deviation of the distribution of each network flux in the 10,000 plausible networks can be seen in Table 2. Given the 100% uncertainty in the low quality parameters, the 95% confidence intervals of nitrification coupled to denitrification and DNRA coupled to anammox did not overlap between the Oligo and Poly sites (Fig. 8). The 95% confidence intervals of nitrification coupled to anammox, however, overlapped by 16% between the two sites. There was little difference between the 95% confidence intervals of N removal capacity at the two sites, which ranged from 3.8% to 5.8% and 2.8% to 6.7% of N input at the Oligo and Poly sites, respectively.

Changing the uncertainty of low quality parameters between ±50% and ±100% of the original network values had little influence on the 95% confidence intervals of the coupling calculation distributions. Doubling the low quality parameter uncertainty generated an increase in the 95% confidence intervals of the distributions of nitrification coupled to denitrification to by 3.9% and 10.1% at the Oligo and Poly sites, respectively. The 95% confidence intervals of the nitrification coupled to anammox incrased by 2.8% and 8.7%, while the intervals of DNRA coupled to anammox increased by 1.6% and 3.3%, at the Oligo and Poly sites, respectively. Thus, the doubling the low quality parameter uncertainty had little impact on our confidence in the analytical results.

**4 Discussion**

Here, we first evaluate the model results in the context of the hypotheses that nitrification coupled to denitrification and anammox would be higher at the Oligo site while DNRA coupled to anammox would be higher at the Poly site. Next, we discuss the implications of the model results for the effects of seawater intrusion on eutrophication and the estuarine N cycle and discuss some of the limitations of the modeling technique used in this work. We conclude by summarizing the contributions of this study to the scientific understanding of process coupling in estuarine N removal.

*4.1 Microbial N process coupling*

The model comparison supports the hypothesis that the coupling of nitrification to denitrification would be higher at the Oligo site than the Poly site. Nitrification coupled to denitrification and was responsible for 43.5% and 18.0% of $N_2$ production through denitrification at the Oligo and Poly sites, respectively. The 95% confidence intervals of nitrification coupled to denitrification distribution produced by the uncertainty analysis did not overlap between the Oligo and Poly sites (Fig. 8), suggesting that this finding is robust given the current knowledge of the CFRE system. Furthermore, Fig. 6 shows that the majority of the reduction in denitrification between the Oligo and Poly sites was a result of decreased coupling to nitrification. This finding is consistent with literature observations, which suggest that denitrification rates in estuaries can be greatly reduced when nitrification is inhibited (An and Joye 2001; Kemp et al. 2005).

The model results also suggested a substantial difference between the coupling of nitrification to anammox at the two sites, as 22.7% and 9.6% of $N_2$ production from anammox was coupled to nitrification at the Oligo and Poly sites, respectively. However, the uncertainty analysis revealed that the 95% confidence intervals of plausible nitrification coupled to anammox estimations overlapped by 16% when using the largest uncertainty for the low quality parameters. This suggests that more precise parameterization data are needed to confirm the hypothesis that nitrification coupled to anammox would be greater at the Oligo site. Specifically, more precise measurements of anammox removal rates may allow the model to confirm a difference between the Oligo and Poly sites, as anammox was the most variable of the direct measurements used for the network parameterization (Table 2).

Despite lower total anammox rates at the Poly site, possibly driven by lower nitrification coupled to anammox, DNRA coupled to anammox increased from 1.8% of anammox $N_2$ production at the Oligo site to 4.8% at the Poly site (Fig. 6). This finding was robust to the uncertainty analysis, as the 95% confidence intervals of the DNRA coupled to anammox distributions did not overlap between the two sites (Fig. 8). The observation that the coupling of DNRA to anammox was stronger at the Poly site is consistent with the findings of previous work, which showed that DNRA plays a relatively minor role in freshwater sediments (Scott et al. 2008). Therefore, the model results suggest that DNRA may play an increasingly important role in estuarine N removal as a result of seawater intrusion.

Although DNRA coupled to anammox was greater at the Poly site, the total increase of $N_2$ produced by DNRA coupled to anammox at the Poly site compared to the Oligo site (0.1 nmol N $cm^{-3}$ $d^{-1}$) was two orders of magnitude smaller than the decrease in $N_2$ produced by nitrification coupled to denitrification and anammox over the same spatial gradient (51.0 nmol N $cm^{-3}$ $d^{-1}$). This result implies that DNRA coupled to anammox alone will not be able to compensate for a reduction in N removal by nitrification coupled to denitrification and anammox caused by saltwater intrusion.

The coupling calculations for the models presented in this work were relatively robust to the uncertainties inherent in the low quality parameters used to construct the network models. The uncertainty analysis showed that at the Oligo site, the 95% confidence intervals of all coupling estimations were altered by less than 3.9% when doubling the uncertainty in low quality parameters (Fig. 8). Coupling calculations at the Poly site were more sensitive to variation in the low quality parameters, with differences in 95% confidence intervals ranging from 3.3% to 10.1% with a doubling of low quality parameter variation. These differences in model calculations were small relative to the magnitude of change in the low quality parameters, highlighting that the low quality parameters used to construct the networks did not have a great impact on the conclusions of this study. For example, although the certainty of the exchange between water column and sediment ammonium (W-NH4 and S-NH4) was low, this process had little impact on the coupling estimations generated by these models.

*4.2 Seawater intrusion and the microbial N cycle*

As seawater intrusion from dredging and sea level rise continues to progress, the environmental conditions for the microbial communities at the Oligo site may shift to more closely resemble the conditions at the Poly site. The model analysis predicts a decoupling of nitrification to denitrification and anammox removal as a result of seawater intrusion, while DNRA coupled to anammox may be enhanced. However, the amount of N removed at both the Oligo and Poly sites was 4.7% and 4.6% of the total N input, respectively (Fig. 7). These findings were robust to the uncertainty analysis and are consistent with reported percentages of denitrification removal in estuaries with similar flushing times, which ranged from 2% to 8% (Nielsen et al. 1995; Nowicki et al. 1997). Estuaries with longer flushing times will likely have higher percentages of N inputs removed through microbial processes (Joye and Anderson 2008).

Although nitrification coupled to denitrification and anammox decreased substantially from the Oligo to the Poly site, direct denitrification increased from 56.5% to 82.0% of denitrification removal. Relative to the N inputs at each site, direct denitrification was able to compensate for reductions in nitrification coupled to denitrification and anammox. The similarity in the percentage of N input converted to $N_2$ gas in each network suggests that seawater intrusion may alter which biogeochemical pathways contributed to N removal, but have little effect on the total amount of $N_2$ produced.

Shifts in which biogeochemical pathways contribute to N removal caused by seawater intrusion may have important implications for the health of estuaries. $NH_4^+$ is converted to $N_2$ gas primarily through nitrification coupled to denitrification. The decoupling of these biogeochemical processes and increased importance of direct denitrification resulting from seawater intrusion, in combination with the decreased adsorption of $NH_4^+$ to sediments (Hou et al. 2003), may decrease $NO_3^-$ pools available to algae while increasing available $NH_4^+$ pools. Under these conditions, the availability of $NO_3^-$ could limit denitrification and anammox $N_2$ production. Because phytoplankton preferentially take up $NH_4^+$ over other forms of inorganic N (McCarthy et al 1977; Carpenter and Dunham 1985), increased $NH_4^+$ pools may lead to larger phytoplankton populations, exacerbating eutrophication.

*4.3 Limitations and future work*

There are several limitations to the modeling techniques used in this study. First, the space-for-time substitution made in this work assumed that the Oligo and Poly sites will behave similarly under equivalent conditions, and cannot address the dynamic processes and transient effects of seawater intrusion (Pickett 1989). The transition from oligohaline to polyhaline conditions may not occur as a smooth interpolation between the Oligo and Poly sites, and may instead pass through alternative transient states that this modeling technique cannot predict. Further, the rate of change in water chemistry may influence how microbial N cycling communities respond to seawater intrusion. Second, this substitution assumed that the Oligo system will not reach an alternative stable state, different from either the Oligo or Poly networks, as a result of the transient dynamics mentioned above. Third, this study presents a comparison of two sites during the summer in a single estuary. While the CFRE is considered to represent a typical coastal plain estuary in the South Eastern United States (Dame et al. 2000), the conclusions of this work may not generalize well to estuaries with different environmental conditions. However, despite the drawbacks inherent in this work, it is a useful first approximation to understanding the potential effect of seawater intrusion on the sedimentary N cycle and provides a basis for future research.

Future studies can build upon the results of the model comparison presented in this work by testing the model predictions using laboratory and field experiments. Mesocosm-scale manipulations of the water chemistry over estuarine sediments in a laboratory can be used to control the environmental conditions proposed to affect microbial communities while carefully monitoring nutrient inputs and transformation process rates. These controlled settings may allow for the collection of more precise measurements of model parameters. It should be noted that the ranges of the 95% confidence intervals of the coupling calculations for nitrification coupled to denitrification and DNRA coupled to anammox, which did not overlap between the Oligo and Poly sites, are separated by a narrow margin and might not be distinct if the models were parameterized with less certainty. More precise parameter measurements will help to ensure that differences between network models can be observed if they are present. Field transplant experiments that exchange the sediments at the Oligo and Poly sites can be used to further examine how microbial process rates in oligohaline sediments respond to seawater intrusion. We hypothesize that the data collected in future experiments will corroborate the conclusions of the model comparison presented in this work by showing shifts in the coupling of microbial processes without shifts in the relative amount of available N removed, and thus will provide further support for the generality of these results.

*4.4 Conclusions*

The model comparison presented here makes five contributions to the scientific understanding of the effects of seawater intrusion on the sedimentary microbial N cycling processes. 1) This work synthesizes disparate measures of nutrient concentrations and microbial transformation processes to generate a whole-system perspective of their interaction. 2) The model results highlight areas where additional measurements are required. 3) This study provides evidence that the reductions in observed rates of $N_2$ production along a salinity gradient in estuarine sediments are likely a result of decreased coupling between nitrification and the removal processes. Modeled DNRA coupled to anammox strengthened at higher salinities, but accounted for less the 5% of anammox $N_2$ production, and therefore did not compensate for reductions in nitrification coupled to denitrification and anammox. 4) The models suggest that seawater intrusion may lead to a higher contribution of direct denitrification as a result of nitrification inhibition, limiting the abilities of estuaries to produce $N_2$ gas from N in the form of $NH_4^+$. Modeled direct denitrification was able to compensate for the reduction in nitrification coupled to denitrification and anammox, as total $N_2$ gas production in relation to the N inputs changed little between the Oligo and Poly sites. However, in estuaries where $NO_3^-$ is not abundant, N removal could become $NO_3^-$ limited as a result of this shift toward a greater reliance on direct denitrification. 5) Our findings imply that seawater intrusion into the freshwater portions of estuaries may exacerbate the effects of nutrient loading and eutrophication through the decreased couplings of nitrification and N removal pathways.  A lessened capacity of estuaries to remove N in the form of $NH_4^+$ could result in longer N residence times and, therefore, in greater resource availability for phytoplankton communities.

**Acknowledgements**

We acknowledge Dr. Michael Mallin and the Lower Cape Fear River Program for contributing data for both the Oligo and Poly sites in the Cape Fear River Estuary, and Dr. Lawrence Cahoon for insight into the Cape Fear River Estuary.  We also acknowledge Leigh Anne Harden for assistance with map creation, as well as anonymous reviewers for helping to improve this work. This work was funded by the US National Science Foundation (DEB1020944) and NSF-OCE (0851435).

**Literature Cited:**
Allesina S, Ulanowicz RE (2004) Cycling in ecological networks: Finn's index revisited. Comput Biol Chem 28: 227-233.
An S, Joye SB (2001) Enhancement of coupled nitrification-denitrification by benthic photosynthesis in shallow estuarine sediments. Limnol Oceanogr 46:62–74.
Anderson DM, Glibert PM, Burkholder JM (2002) Harmful algal blooms and eutrophication: nutrient sources, composition, and consequences. Estuaries 25:704–726.
Baird D, McGlade, JM, Ulanowicz, RE (1991) The comparative ecology of six marine ecosystems. Philos Trans R Soc Lond B. 33:15–29
Baird D, Heymans JJ (1996) Assessment of ecosystem changes in response to freshwater inflow of the Kromme River Estuary, St. Francis Bay, South Africa: A network analysis approach. Water SA 22:307–318.
Barbier EB, Hacker SD, Kennedy C, Koch EW, Stier AC, Silliman BR (2011) The value of estuarine and coastal ecosystem services. Ecol Monogr 81:169–193.
Baron JS, Hall EK, Nolan BT, Finlay JC, Bernhardt ES, Harrison JA, Chan F, Boyer EW (2013) The interactive effects of excess reactive nitrogen and climate change on aquatic ecosystems and water resources of the United States. Biogeochemistry 114:71-92. doi: 10.1007/s10533-12-9788-y
Bernhard AE, Tucker J, Giblin AE, Stahl DA (2007). Functionally distinct communities of ammonia-oxidizing bacteria along an estuarine salinity gradient. Environ Microbiol 9:1439–1447.
Berounsky VM, Nixon SW (1993) Rates of nitrification along and estuarine gradient in Narragansett Bay. Estuaries 16:718–730.
Blackburn TH (1988) Benthic mineralization and bacterial production, 175-190. In: Blackburn TH, Sørensen J (eds.) Nitrogen cycling in coastal marine environments, John Wiley and Sons.
Borrett SR, Freeze MA (2011) Reconnecting environs to their environment. Ecol Model 222:2393–2403. doi:10.1016/j.ecolmodel.2010.10.015
Borrett SR, Osidele, OO (2007) Environ indicator sensitivity to flux uncertainty in a phosphorus model of Lake Sidney Lanier, USA. Ecol Model 200:371–383.
Borrett SR, Whipple SJ, Patten BC (2010) Rapid development of indirect effects in ecological networks. Oikos 119:1136–1148.
Bouvier TC, del Giorgio PA (2002) Compositional changes in free-living bacterial communities along a salinity gradient in two temperate estuaries. Limnol Oceanogr 47:453–470.

Carpenter EJ, Dunham S (1985) Nitrogenous nutrient uptake, primary production, and species composition of phytoplankton in the Carmas River estuary, Long Island, New York. Limnol Oceanogr 30:513-526.

Christian RR, Thomas CR (2003) Network analysis of nitrogen inputs and cycling in the Neuse River Estuary, North Carolina, USA. Estuaries 26:815–828.

Christian RR, Baird D, Luczkovich J, Johnson JC, Scharler UM, Ulanowicz RE (2005) Role of network analysis in comparative ecosystem ecology of estuaries. In: Belgrano A (ed), Aquatic food webs: An ecosystem approach, Oxford University Press, pp 25-40.

Costanza R (1992) Toward an operational definition of ecosystem health. In: Costanza R, Norton BG, Haskell BD (eds.) Ecosystem health: New goals for environmental management, Island Press, pp 239-256.

Costanza R, d'Arge R, De Groot R, Farber S, Grasso M, Hannon B, Limburg K, Naeem S, O'Neill RV, Paruelo J, Raskin RG, Sutton P, and van den Belt M (1998) The value of the world's ecosystem services and natural capital. Nature 387:253–260.

Cowan JLW, Pennock JR, Boynton WR (1996) Seasonal and interannual patterns of sediment-water nutrient and oxygen fluxes in Mobile, Alabama (USA): Regulating factors and ecological significance. Mar Ecol Prog Ser 141:229–245.

Craft C, Clough J, Ehman J, Joye SB, Park R, Pennings S, Guo H, and Machmuller M (2008) Forecasting the effects of accelerated sea-level rise on tidal marsh ecosystem services. Front Ecol Environ 7:73–78.

Dame R, Alber M, Allen D, Mallin M, Montague C, Lewitus A, Chalmers A, Gardner R, Gilman C, Kjerfve B, Pinckney J, Smith N (2000) Estuaries of the south Atlantic coast of North America: Their geographical signatures. Estuaries 23:793–819.

Dong L, Thornton D, Nedwell D, Underwood G (2000) Denitrification in sediments of the River Colne estuary, England. Mar Ecol Prog Ser 203:109–122.

Ensign SH, Halls JN, Mallin MA (2004) Application of digital bathymetry data in an analysis of flushing times of two North Carolina estuaries. Comput Geos 30:501–511.

Fath BD, Patten BC (1999) Review of the foundations of network environ analysis. Ecosystems 2:167–179.

Fath BD, Scharler UM, Ulanowicz RE, and Hannon B (2007) Ecological network analysis: Network construction. Ecol Model 208:49-55.

Finn JT (1980) Flow analysis of models of the Hubbard Brook ecosystem. Ecology 61:562–571.

Francis CA, Beman JM, Kuypers MM (2007) New processes and players in the nitrogen cycle: The microbial ecology of anaerobic and archaeal ammonia oxidation. ISME J 1:19–27.

Gardner WS, Seitzinger SP, Malczyk JM (1991) The effects of sea salts on the forms of nitrogen released from estuarine and freshwater sediments: Does ion pairing affect ammonium flux? Estuaries 14:157–166.

Giblin AE, Weston NB, Banta GT, Tucker J, Hopkinson CS (2010) The effects of salinity on nitrogen losses from an oligohaline estuarine sediment. Estuar Coast 33:1054–1068.

Graham TB (2008) Nitrate recycling versus removal in the Cape Fear River Estuary. Master's

thesis, University of North Carolina Wilmington.
Grant J, Cranford P, Emerson C (1997) Sediment resuspension rates, organic matter quality and food utilization by sea scallops (*Placopecten magellanicus*) on Georges Bank. J Mar Res 55:965–994.
Hackney CT, Yelverton GF (1990) Effects of human activities and sea level rise on wetland ecosystems in the Cape Fear River Estuary, North Carolina, USA. In: Whigham DF (ed.) Wetland ecology and management: Case studies, Springer, pp 55-61.
Hannon B (1973) The structure of ecosystems. J Theor Biol 41:535–546.
Hannon B (1985) Ecosystem flow analysis. In: Ulanowicz RE, Platt T (eds.) Ecosystem theory for biological oceanography, Can Bul Fish Aquat Sci, pp 97 118
Hansen JI, Henriksen K, Blackburn TH (1981) Seasonal distribution of nitrifying bacteria and rates of nitrification in coastal marine sediments. Microb Ecol 7:297–304.
Henriksen K, Kemp WM (1988) Nitrification in estuarine and coastal marine sediments: Methods, patterns and regulating factors. In: Blackburn TH and Sørensen J (eds.) Nitrogen cycling in coastal marine environments, John Wiley and Sons, pp 207-250.
Heymans JJ, Ulanowicz RE, Bondavalli C (2002) Network analysis of the South Florida Everglades graminoid marshes and comparison with nearby cypress ecosystems. Ecol Model 149:5–23.
Hines DE, Lisa JA, Song B, Tobias CR, Borrett SR (2012). A network model shows the importance of coupled processes in the microbial N cycle in the Cape Fear River Estuary. Estuar Coast Shelf S 106:45-57.
Hirsch MD (2010) Anammox and denitrification in the Cape Fear River Estuary: anammox bacterial diversity and significant in sedimentary nitrogen removal. Master's thesis, University of North Carolina Wilmington.
Hou L, Liu M, Jiang H, Xu S, Ou D, Liu Q, Zhang B (2003) Ammonium adsorption by tidal flat surface sediments from the Yangtze Estuary. Environ Geol 45:72–78.
Howarth RW, Marino R (2006) Nitrogen as the limiting nutrient for eutrophication in coastal marine ecosystems: evolving views over three decades. Limnol Oceanogr 51:364–376.
IPCC (2007) Intergovernmental Panel on Climate Change Assessment Report 4. www.ipcc.ch/publications_and_data/ar4/syr/en/main.html
Jenkins MC, Kemp WM (1984) The coupling of nitrification and denitrification in two estuarine sediments. Limnol Oceanogr 29:609–619.
Jordan TE, Correll DL, Whigham DF (1983) Nutrient flux in the Rhode River: Tidal exchange of nutrients by brackish marshes. Estuar Coast Shelf S 17:651–667.
Joye SB, Anderson IC (2008) Nitrogen cycling in coastal sediments, 868-900. In: Capone DG (ed.), Nitrogen in the marine environment, 2$^{nd}$ edn. Elsevier, p 868-900.
Joye SB, Hollibaugh JT (1995) Influence of sulfide inhibition of nitrification on nitrogen regeneration in sediments. Science 270:623–625.
Kones, JK, Soetaert K, van Oevelen D, Owino JO (2009) Are network indecies robust indicators of food web function? A Monte Carlo approach. Ecol Model 220:370-382.
Kemp WM, Boynton WR, Adolf JE, Boesch DF, Boicourt WC, Brush G, Cornwell JC, Fisher TR, Glibert PM, Hagy JD, Harding LW, Houde ED, Kimmel DG, Miller WD, Newell RIE, Roman MR, Smith EM, Stevenson JC (2005). Eutrophication of Chesapeake Bay: Historical trends and ecological interactions. Mar Ecol Prog Ser 303:1–29.

Kemp WM, Sampou P, Caffrey J, Mayer M, Henriksen K, Boynton WR (1990) Ammonium recycling versus denitrification in Chesapeake Bay sediments. Limnol Oceanogr 35:1545–1563.

Lau MK, Borrett SR, Hines DE (2012) enaR: Tools ecological network analysis. R package version 1.01, cran.r-project.org/web/packages/enaR/index.html

Mallin MA, McIver MR, Merrit JF (2009) Environmental assessment of the lower Cape Fear River system, 2008. Tech. Rep. 09-06, University of North Carolina Wilmington, uncw.edu/cms/aelab/LCFRP/

Mallin MA, McIver MR, Merrit JF (2010) Environmental assessment of the lower Cape Fear River system, 2009. Tech. Rep. 10-04, University of North Carolina Wilmington, uncw.edu/cms/aelab/LCFRP/

McCarthy JJ, Taylor WR, Taft JL (1977) Nitrogenous nutrition of the plankton in the Chesapeake Bay 1: Nutrient availability and phytoplankton preferences. Limnol Oceanogr 22:996-1011.

Nielsen K, Nielsen LP, Rasmussen P (1995) Estuarine nitrogen retention independently estimated by the denitrification rate and mass balance methods: A study of Norsminde Fjord, Denmark. Mar Ecol Prog Ser 119:275–283.

Newport BD (1977) Salt water intrusion in the United States. Environmental Protection Agency, Office of Research and Development, Robert S. Kerr Environmental Research Laboratory.

Nowicki, BL, Kelly JR, Requintina E, van Keuren D (1997) Nitrogen losses through sediment denitrification in Boston Harbor and Massachusetts Bay. Estuaries 20:626–639.

Oreskes, N, Shrader-Frechette, K, Belitz, K (1994) Verification, validation, and confirmation of numerical-models in the earth sciences. Science 263:641–646.

Patten BC (1978) Systems approach to the concept of environment. Ohio J Sci 78:206–222.

Patten BC (1981) Environs: The superniches of ecosystems. Am Zool 21:845-852.

Patten BC (1982) Environs: Relativistic elementary particles for ecology. Am Nat 119:179–219.

Patten BC, Bosserman RW, Finn JT, Cale WG (1976) Propagation of cause in ecosystems. In: Patten BC (ed.) Systems analysis and simulation in ecology, Vol. IV, Academic Press, pp 457-579.

Patten BC, Higashi, M, Burns TP (1990) Trophic dynamics in ecosystem networks: significance of cycles and storage. Ecol Model 51:1–28.

Pickett ST (1989) Space-for-time substitution as an alternative to long-term studies. In: Likens GE (ed.) Long-term studies in ecology, Springer, pp 110-135.

Pinckney JL, Paerl HW, Tester P, Richardson TL (2001) The role of nutrient loading and eutrophication in estuarine ecology. Environ Health Persp 109:699-706.

Pujo-Pay M, Conan P, Raimbault P (1997) Secretion of dissolved organic nitrogen by phytoplankton assessed by wet oxidation and $^{15}N$ tracer procedures. Mar Ecol Prog Ser 153:99–111.

Rysgaard S, Thastum P, Dalsgaard T, Christensen PB, Sloth NP (1999) Effects of salinity on $NH_4^+$ adsorption, nitrification, and denitrification in Danish estuarine sediments. Estuaries 22:21–30.

Ryther JH, Dunstan WM (1971) Nitrogen, phosphorus, and eutrophication in the coastal marine environment. Science 171:1008–1013.

Santoro AE (2010) Microbial nitrogen cycling at the saltwater-freshwater interface. Hydrogeol J 18:187–202.

Salas, AK, Borrett SR (2011) Evidence for dominance of indirect effects in 50 trophic ecosystem networks. Ecol Model 222:1192–1204.

Schramski JR, Kazanci C, Tollner EW (2011) Network environ theory, simulation and Econet © 2.0. Environ Model Softw 26:419–428. doi:10.1016/j.envsoft.2010.10.003

Scott TJ, McCarth MJ, Gardner WS, Doyle RD (2008) Denitrification, dissimilatory nitrate reduction to ammonium, and nitrogen fixation along a nitrate concentration gradient in a created freshwater wetland. Biogeochemistry 87:99-111. doi 10.1007/s10533-007-9171-6

Seitzinger S, Harrison J, Böhlke J, Bouwman A, Lowrance R, Peterson B, Tobias C, and Drecht G (2006) Denitrification across landscapes and waterscapes: a synthesis. Ecol Appl 16:2064–2090.

Soetaert K, Van den Meersche K, van Oevelen D (2009) limSolve: Solving linear inverse models. R package version 1.5.1, cran.r-project.org/web/packages/limSolve/index.html

Thamdrup B, Dalsgaard T (2002) Production of $N_2$ through anaerobic ammonium oxidation coupled to nitrate reduction in marine sediments. Appl Environ Microb 68:1312–1318.

Tobias CR, Harvey JW, Anderson IC (2001) Quantifying groundwater discharge through fringing wetlands to estuaries: Seasonal variability, method comparison, and implications for wetland-estuary exchange. Limnol Oceanogr 46:604–615.

Ulanowicz RE (1986) Growth and development: Ecosystems phenomenology. Springer–Verlag.

Ulanowicz RE (2004) Quantitative methods for ecological network analysis. Comput Biol Chem 28:321–339. dx.doi.org/10.1016/j.compbiolchem.2004.09.001

Ulanowicz RE, Kay J (1991). A package for the analysis of ecosystem flow networks. Environ Softw 6:131–142.

Veuger D, Middleburg JJ, Boschker HTS, Nieuwenhuize J, van Rijswijk P, Rochelle-Newall EJ, Navarro N (2004) Microbial uptake of dissolved organic and inorganic nitrogen in Randers Fjord. Estuar Coast Shelf S 61:507–515.

Whipple SJ, Borrett SR, Patten BC, Gattie DK, Schramski JR, Bata SA (2007) Indirect effects and distributed control in ecosystems: Comparative network environ analysis of a seven-compartment model of nitrogen flow in the Neuse River Estuary, USA—Time series analysis. Ecol Model 206:1–17

Whitman WB, Coleman DC, Wiebe WJ (1998) Prokaryotes: The unseen majority. Proc Nat Acad Sci USA 95:6578–6583.

Zhang Z, Cui B, Ou B, Fan X (2012) Wetland network design for mitigation of saltwater intrusion by transferring tidal discharge. Clean–Soil, Air, Water 40:1057–1063.

**Table 1:** Fluxes, parameter values, and sources for the oligohaline (Oligo) and polyhaline (Poly) networks. Boundary flows represent network inputs and outputs, while internal fluxes represent flows from one compartment to another. Parameter values are in nmol N cm$^{-3}$ d$^{-1}$. Sources used only for the Oligo network are denoted by †, while sources used only for the Poly network are denoted by ‡. Unmarked sources were used for both networks. Values for the Oligo model were previously presented in Hines et al. (2012).

| Flux | Oligo | Poly | Source |
| --- | --- | --- | --- |
| boundary → W-NH$_4$ | 130.0 | 72.5 | direct measurements- Ensign et al. 2004; Hirsch 2010 |
| boundary → W-NO$_X$ | 1020.0 | 381.1 | direct measurements- Ensign et al. 2004; Hirsch 2010 |
| boundary → W-M | 3.9x10$^{-5}$ | 3.9x10$^{-5}$ | Whitman et al. 1998 |
| boundary → W-ON | 1160.0 | 1255.1 | direct measurements- Ensign et al. 2004; Mallin et al. 2010 |
| boundary → S-NH$_4$ | 1238.2 | 975.1 | mass balance |
| boundary → S-NO$_X$ | 173.2 | 345.5 | direct measurments- Ensign et al. 2004; Hirsch 2010 |
| boundary → S-ON | 79.0 | 39.1 | Jordan et al. 1983 |
| W-NH$_4$ → boundary | 276.0 | 132.4 | direct measurements- Ensign et al. 2004; Hirsch 2010 |
| W-NO$_X$ → boundary | 1008.6 | 380.9 | direct measurements- Ensign et al. 2004; Hirsch 2010 |
| W-M → boundary | 3.9x10$^{-5}$ | 3.9x10$^{-5}$ | Whitman et al. 1998 |
| W-ON → boundary | 1159.6 | 1246.8 | direct measurements- Ensign et al. 2004; Mallin et al. 2010 |
| S-NH$_4$ → boundary | 1080.0 | 1006.9 | Tobias et al. 2001 |
| S-NO$_X$ → boundary | 6.0 | 127.3 | Tobias et al. 2001 |
| S-ON → boundary | 104.1 | 32.7 | Jordan et al. 1983 |
| S-NH$_4$ anammox | 2.5 | 1.8 | direct measurements- Hirsch 2010 |
| S-NO$_X$ anammox | 2.5 | 1.8 | direct measurements- Hirsch 2010 |
| S-NO$_X$ denitrification | 172.0 | 136.7 | direct measurements- Hirsch 2010 |
| S-NO$_X$ burial | 0.3 | 7.8x10$^{-3}$ | estimation from sea level rise |
| S-M burial | 3.9x10$^{-7}$ | 3.9x10$^{-7}$ | estimation from sea level rise |
| S-ON burial | 3.9 | 2.0 | estimation from sea level rise |
| W-NH$_4$ → W-NO$_X$ | 1.7 | 0.7 | Kemp et al. 1990†; Berounsky and Nixon 1993†; Whitman et al. 1998‡ |
| W-NH$_4$ → W-M | 1.9 | 1.7 | Veuger et al. 2004 |
| W-NH$_4$ → S-NH$_4$ | 5.5 | 1.5 | Cowan et al. 1996 |
| W-NO$_X$ → W-M | 9.8 | 0.5 | Veuger et al. 2004 |
| W-NO$_X$ → S-NO$_X$ | 14.1 | 7.7 | Cowan et al. 1996 |
| W-M → W-NH$_4$ | 3.1 | 3.1 | mass balance |
| W-M → W-ON | 16.0 | 0.5 | mass balance |
| W-M → S-M | 119.2 | 119.2 | Cowan et al. 1996 |
| W-ON → W-NH$_4$ | 5.2 | 5.4 | Pujo-Pay et al. 1997 |
| W-ON → W-M | 7.4 | 1.4 | Veuger et al. 2004 |
| W-ON → S-ON | 853.9 | 425.8 | estimation from sea level rise |
| S-NH$_4$ → S-NO$_X$ | 144.0 | 77.5 | Hansen et al. 1981‡; Henriksen and Kemp 1988†; Kemp et al. 1990† |
| S-NH$_4$ → S-M | 212.8 | 186.2 | Whitman et al. 1998‡; Veuger et al. 2004† |
| S-NH$_4$ → W-NH$_4$ | 136.5 | 55.3 | mass balance |
| S-NO$_X$ → S-NH$_4$ | 39.0 | 104.4 | direct measurements- Graham 2008 |
| S-NO$_X$ → S-M | 109.0 | 53.2 | Whitman et al. 1998‡; Veuger et al. 2004† |
| S-NO$_X$ → W-NO$_X$ | 2.1 | 7.3 | Cowan et al. 1996 |
| S-M → S-NH$_4$ | 146.7 | 146.7 | mass balance |
| S-M → S-ON | 257.1 | 253.2 | mass balance |
| S-M → W-M | 119.2 | 119.2 | Cowan et al. 1996 |
| S-ON → S-NH$_4$ | 150.0 | 100.0 | Blackburn 1988 |
| S-ON → S-M | 82.0 | 159.6 | Whitman et al. 1998‡; Veuger et al. 2004† |
| S-ON → W-ON | 850.0 | 423.8 | Grant et al. 1997 |

**Table 2:** Network fluxes by parameter quality according to the Costanza (1992) rubric; (H) high, (M) medium, (L) low. % disturbance shows the restriction range above and below the original network values used in the whole network uncertainty analysis. Mean and standard deviation (SD) are shown for the distributions of parameter values observed in the plausible networks at the Oligo and Poly sites. Parameter values have units of nmol N cm$^{-3}$ d$^{-1}$.

| Flux | Quality | Oligo % Disturbance | Oligo Mean | Oligo SD | Poly % Disturbance | Poly Mean | Poly SD |
|---|---|---|---|---|---|---|---|
| boundary → W-NH$_4$ | H | 40 | 129.8 | 29.8 | 41 | 72.7 | 16.8 |
| boundary → W-NO$_X$ | H | 32 | 1020.3 | 185.2 | 37 | 379.1 | 80.8 |
| boundary → W-ON | H | 47 | 1158.1 | 210.9 | 28 | 1254.9 | 114.8 |
| boundary → S-NO$_X$ | H | 16 | 172.9 | 15.8 | 43 | 345.5 | 55.0 |
| W-NH$_4$ → boundary | H | 38 | 276.2 | 60.5 | 17 | 131.7 | 12.4 |
| W-NO$_X$ → boundary | H | 37 | 1010.3 | 187.0 | 38 | 379.0 | 80.8 |
| W-ON → boundary | H | 28 | 1158.5 | 186.4 | 11 | 1246.9 | 77.2 |
| S-NH$_4$ anammox | H | 47 | 2.5 | 0.6 | 36 | 1.8 | 0.4 |
| S-NO$_X$ anammox | H | 47 | 2.5 | 0.6 | 36 | 1.8 | 0.4 |
| S-NO$_X$ denitrification | H | 9 | 171.2 | 9.5 | 42 | 136.6 | 31.8 |
| S-NO$_X$ → S-NH$_4$ | H | 23 | 38.9 | 5.1 | 19 | 104.4 | 11.4 |
| boundary → W-M | M | 47 | 3.9x10$^{-5}$ | 1.1x10$^{-5}$ | 47 | 4.0x10$^{-5}$ | 1.1x10$^{-5}$ |
| boundary → S-ON | M | 47 | 78.9 | 21.3 | 47 | 39.1 | 9.5 |
| W-M → boundary | M | 47 | 3.0x10$^{-5}$ | 1.1x10$^{-5}$ | 47 | 4.0x10$^{-5}$ | 1.1x10$^{-5}$ |
| S-NH$_4$ → boundary | M | 47 | 1081.1 | 266.1 | 47 | 1013.9 | 251.8 |
| S-NO$_X$ → boundary | M | 47 | 6.0 | 1.6 | 47 | 127.1 | 34.6 |
| S-ON → boundary | M | 47 | 104.2 | 28.1 | 47 | 33.0 | 8.0 |
| W-NH$_4$ → W-NO$_X$ | M | 47 | 1.7 | 0.5 | 47 | 0.7 | 0.2 |
| W-NH$_4$ → W-M | M | 47 | 1.9 | 0.5 | 47 | 1.8 | 0.5 |
| W-NH$_4$ → S-NH$_4$ | M | 47 | 5.5 | 1.5 | 47 | 1.5 | 0.4 |
| W-NO$_X$ → W-M | M | 47 | 9.8 | 2.7 | 47 | 0.5 | 0.1 |
| W-NO$_X$ → S-NO$_X$ | M | 47 | 14.1 | 3.7 | 47 | 7.7 | 2.1 |
| W-M → S-M | M | 47 | 118.9 | 29.4 | 47 | 119.5 | 31.3 |
| W-ON → W-NH$_4$ | M | 47 | 5.2 | 1.4 | 47 | 5.4 | 1.5 |
| W-ON → W-M | M | 47 | 7.4 | 2.0 | 47 | 1.4 | 0.4 |
| S-NH$_4$ → S-NO$_X$ | M | 47 | 143.4 | 32.6 | 47 | 77.1 | 21.1 |
| S-NH$_4$ → S-M | M | 47 | 213.2 | 57.4 | 47 | 186.9 | 47.0 |
| S-NO$_X$ → S-M | M | 47 | 109.0 | 28.9 | 47 | 53.1 | 14.4 |
| S-NO$_X$ → W-NO$_X$ | M | 47 | 2.2 | 0.6 | 47 | 7.3 | 1.9 |
| S-M → W-M | M | 47 | 118.9 | 29.3 | 47 | 119.5 | 31.2 |
| S-ON → S-NH$_4$ | M | 47 | 149.9 | 40.9 | 47 | 99.9 | 23.8 |
| S-ON → S-M | M | 47 | 82.3 | 20.0 | 47 | 159.2 | 39.9 |
| S-ON → W-ON | M | 47 | 851.4 | 230.1 | 47 | 422.6 | 114.8 |
| boundary → S-NH$_4$ | L | 100 | 1241.5 | 283.0 | 100 | 982.3 | 260.2 |
| S-NO$_X$ burial | L | 100 | 0.3 | 0.2 | 100 | 7.8x10$^{-3}$ | 4.5x10$^{-3}$ |
| S-M burial | L | 100 | 3.9x10$^{-7}$ | 2.2x10$^{-7}$ | 100 | 4.0x10$^{-7}$ | 2.6x10$^{-7}$ |
| S-ON burial | L | 100 | 3.9 | 2.2 | 100 | 2.0 | 1.2 |
| W-M → W-NH$_4$ | L | 100 | 3.1 | 1.8 | 100 | 3.1 | 1.8 |
| W-M → W-ON | L | 100 | 15.9 | 8.2 | 100 | 0.5 | 0.3 |
| W-ON → S-ON | L | 100 | 854.9 | 254.3 | 100 | 424.4 | 128.3 |
| S-NH$_4$ → W-NH$_4$ | L | 100 | 137.3 | 69.2 | 100 | 54.5 | 20.7 |
| S-M → S-NH$_4$ | L | 100 | 146.6 | 83.5 | 100 | 146.0 | 76.1 |
| S-M → S-ON | L | 100 | 257.9 | 99.2 | 100 | 253.2 | 89.1 |

**Figure Legends**

**Figure 1:** Map of the Cape Fear River Estuary, North Carolina. Horseshoe Bend (Oligo) and Marker 35 (Poly) study sites marked by arrows. City of Wilmington shown by black circle.

**Figure 2:** Network models constructed at the (A) oligohaline and (B) polyhaline sites. Network structure was identical between the two sites, while flux magnitudes varied. Arrow widths approximate relative flux magnitudes within each network. Labeled loss arrows represent (a) anammox removal, (b) denitrification removal, (c) nitrate and nitrite burial, (d) microbial burial, and (e) organic N burial. Italicized numbers in the bottom left of node boxes represent standing stock concentrations, while underlined numbers in the top left of node boxes show the node label number.

**Figure 3:** Example calculation of coupling in the denitrification environ for percent nitrification coupled to denitrification. (A) Denitrification, (B) microbial uptake of $NO_x$ in the sediments, (C) transfer of $NO_x$ from the sediments to the water column, (D) DNRA, and (X) nitrification.

**Figure 4:** Realized input environs for the denitrification pathway at (A) oligohaline and (B) polyhaline sites. Network outputs highlighted in gray. Arrow widths approximate magnitudes of fluxes in each realized environ.

**Figure 5:** Realized input environs for anammox $N_2$ production at (A) oligohaline and (B) polyhaline sites. Network outputs highlighted in gray. Arrow widths approximate magnitudes of fluxes in each realized environ.

**Figure 6:** Estimated coupling of N transformation and removal processes at oligohaline and polyhaline sites for (A, B) denitrification and (C, D) anammox in (A, C) absolute units and (B, D) as a percentage of each process. Abbreviations: direct denitrification (direct DNT), nitrification coupled to denitrification (NTR - DNT), direct anammox (direct AMX), DNRA coupled to anammox (DNRA- AMX), nitrification coupled to anammox (NTR - AMX), oligohaline (O), polyhaline (P).

**Figure 7:** Total the amount of direct and coupled microbial processes involved in (A) $N_2$ production and (B) the proportion of N involved in removal processes relative to the N input at each site.

**Figure 8:** Uncertainty analysis showing the 95% confidence intervals of estimations for nitrification coupled to denitrification, nitrification coupled to anammox, and DNRA coupled to anammox based on 10,000 plausible network parameterizations. Large dots indicate the values calculated by the original network parameterizations, while solid horizontal lines indicate the 95% range of each coupling calculation distribution. Vertical lines denote the 95% confidence intervals for estimates with ±50% variation in low quality parameters. Horizontal lines that extend past the vertical lines show the 95% confidence intervals for estimates with ±100% variation in low quality parameters.

**Figures**

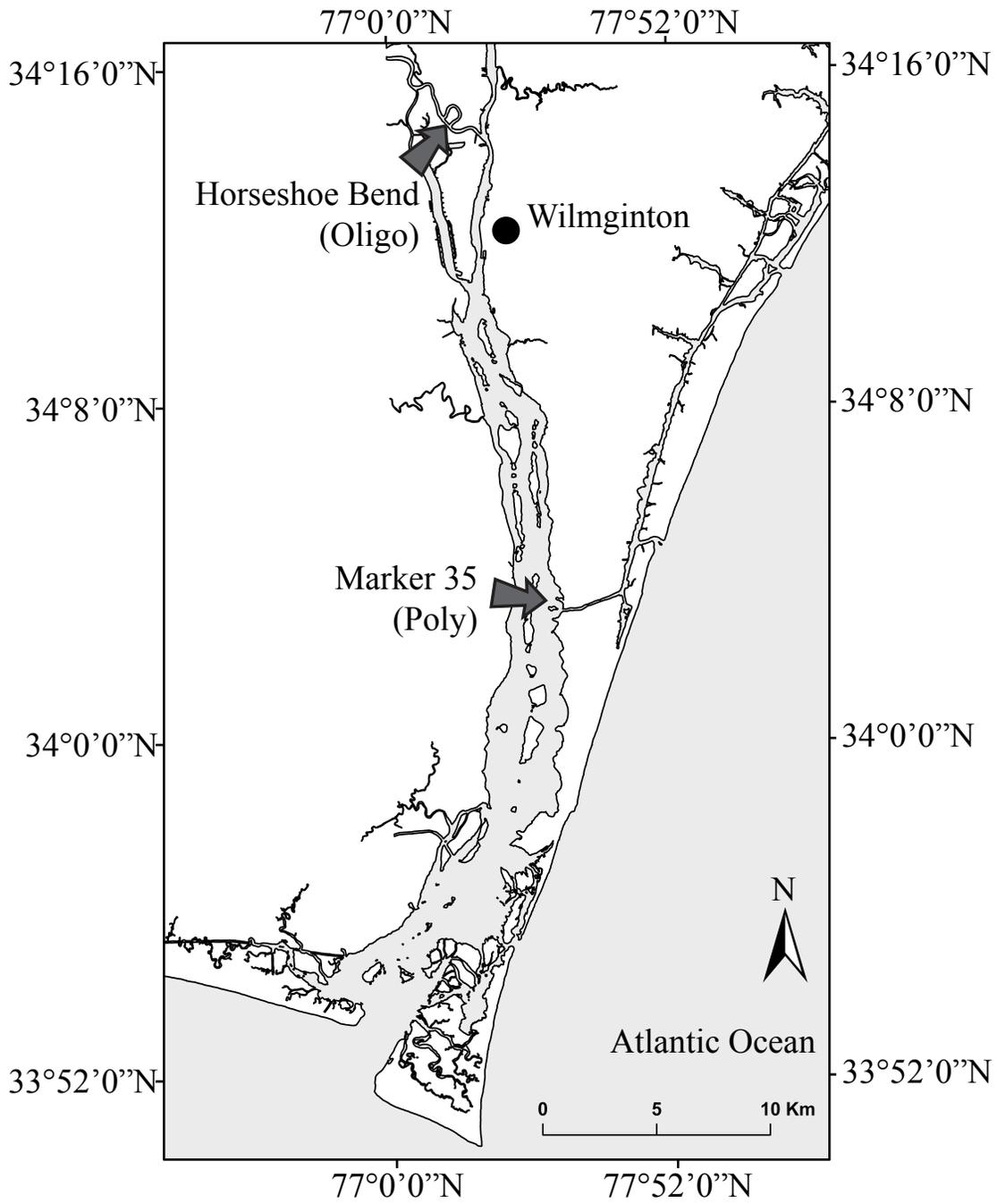

**Figure 1**

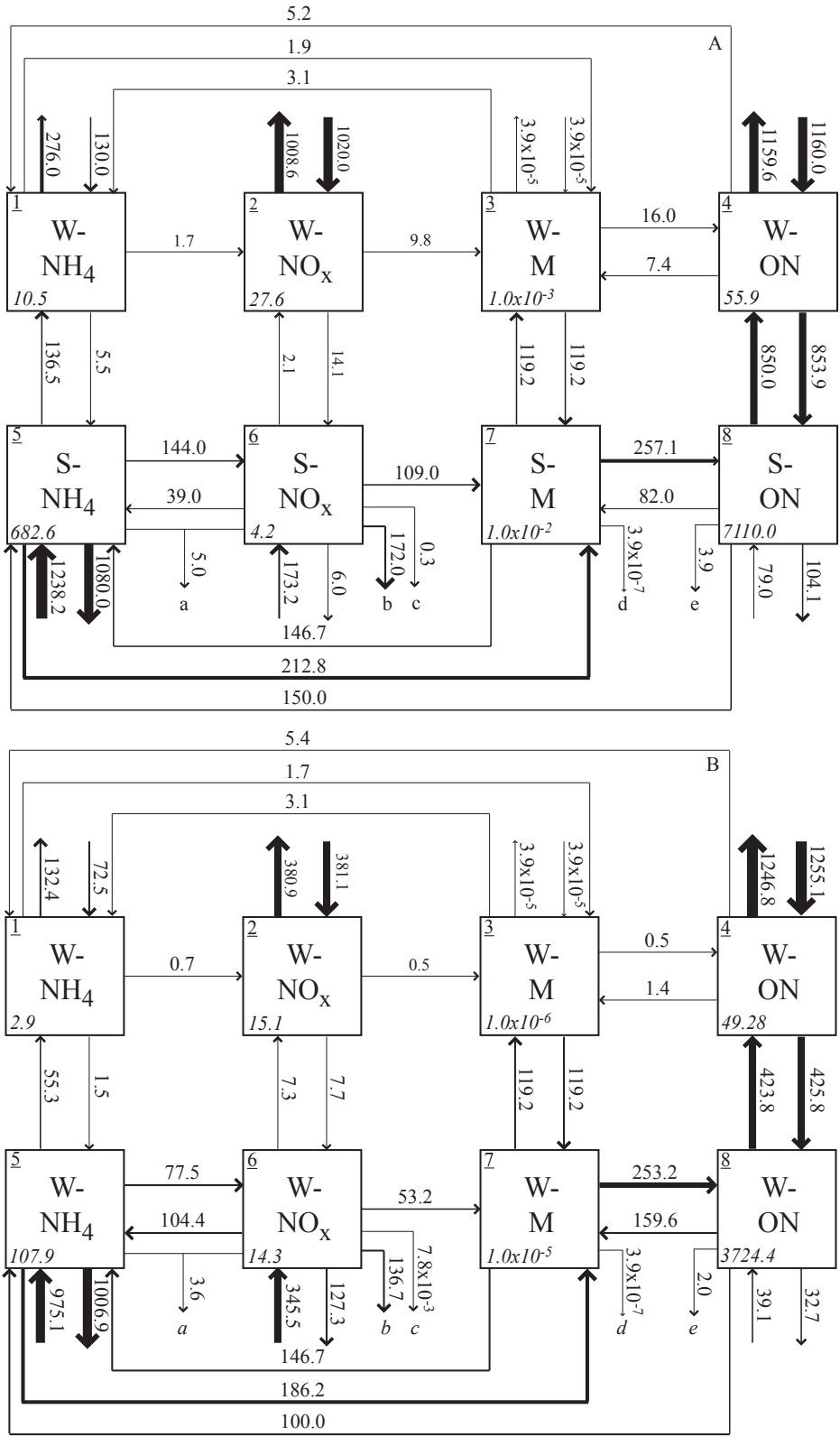

**Figure 2**

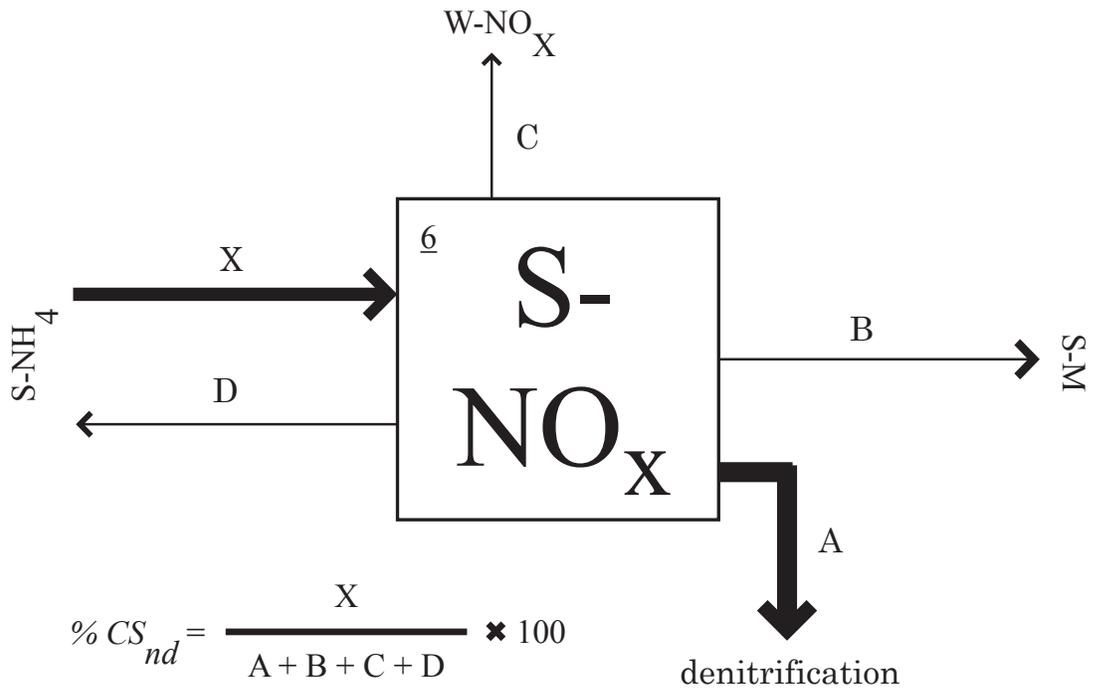

$$\% CS_{nd} = \frac{X}{A + B + C + D} \times 100$$

**Figure 3**

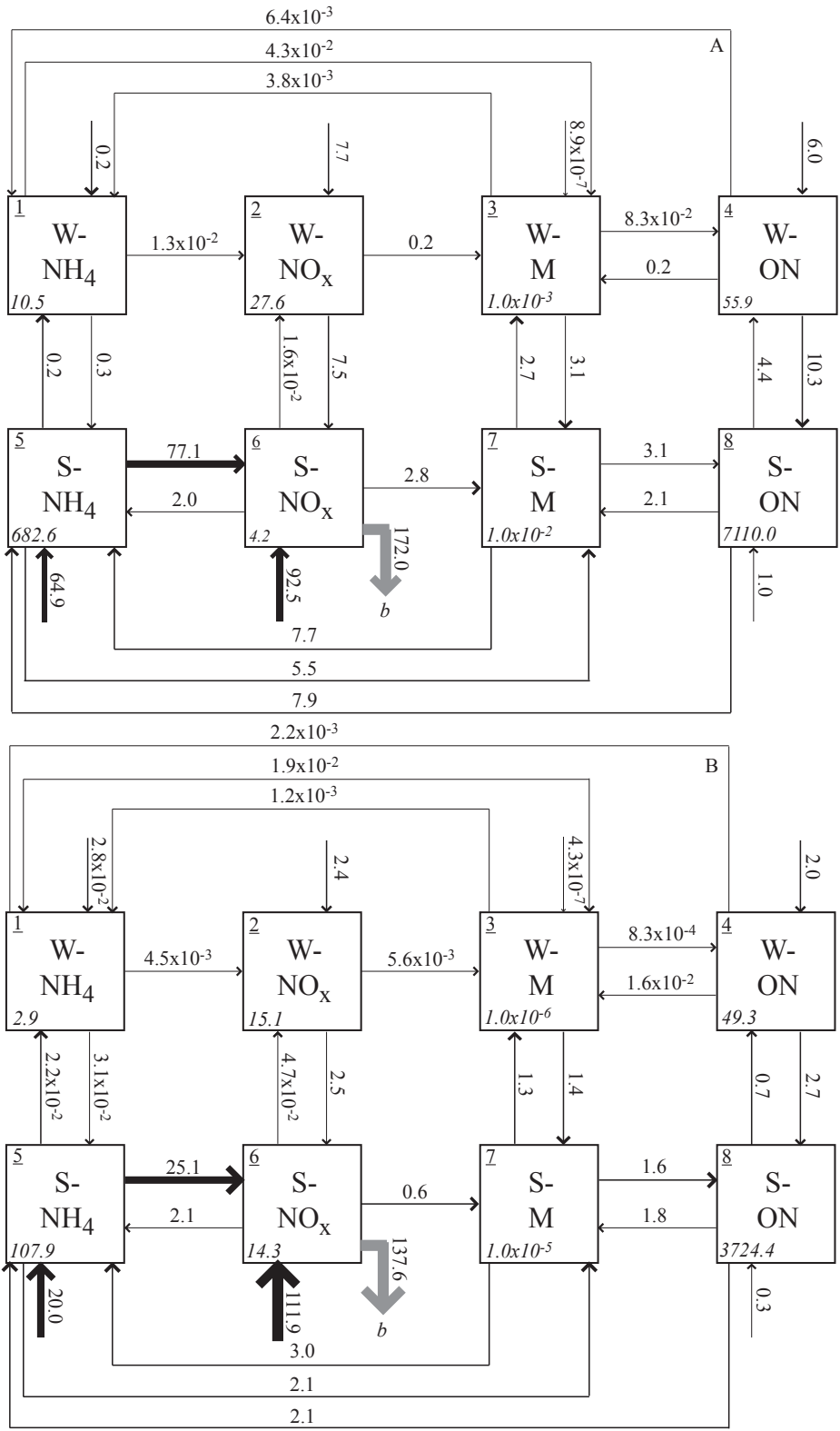

**Figure 4**

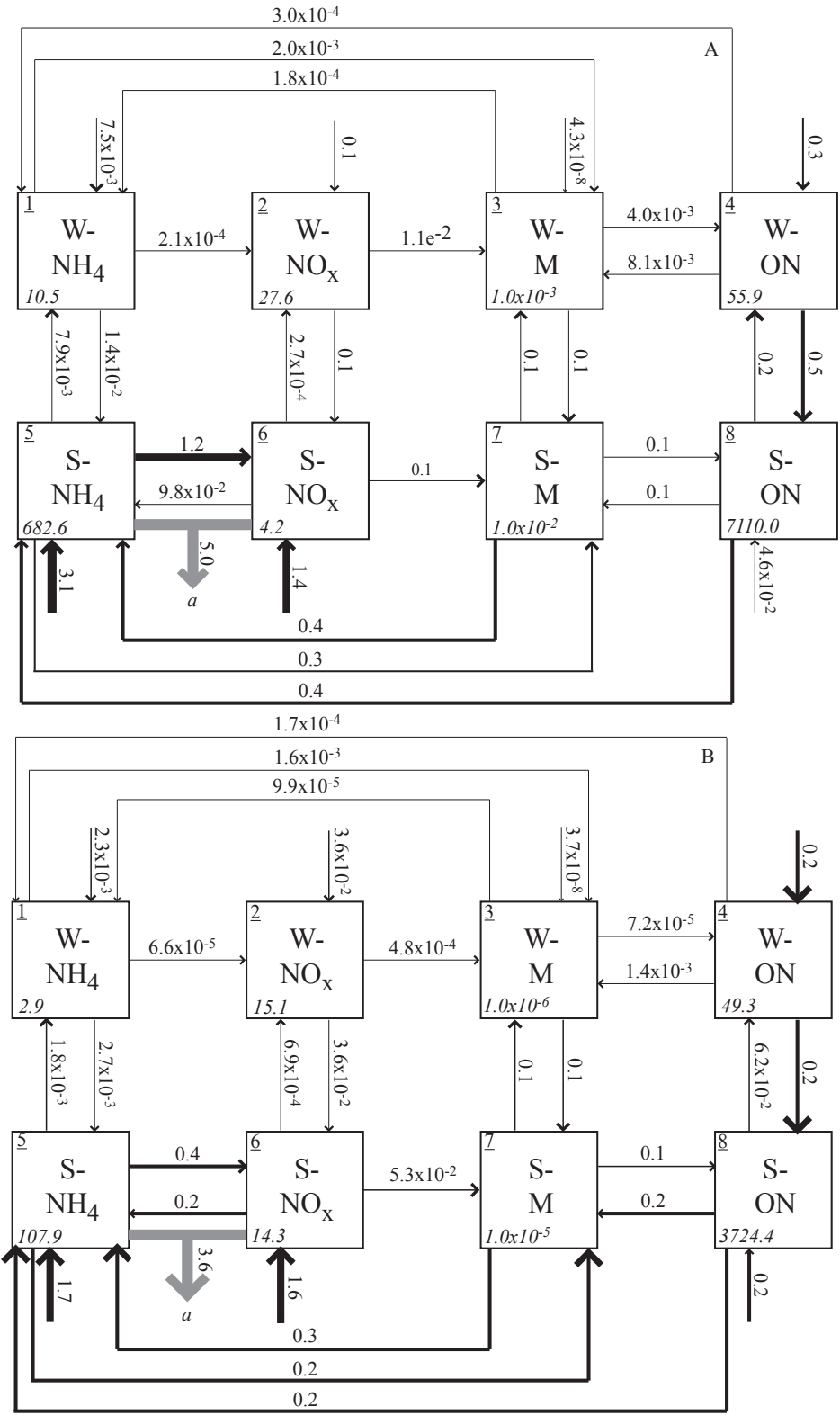

**Figure 5**

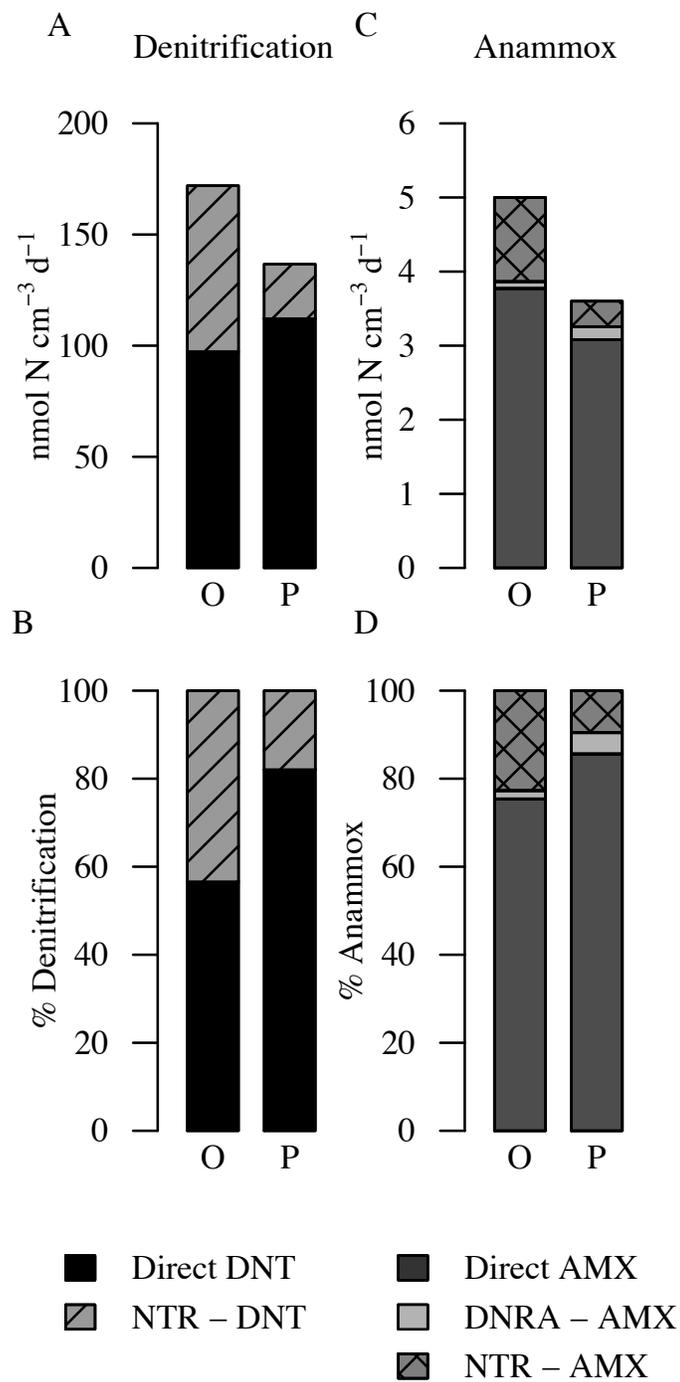

Figure 6

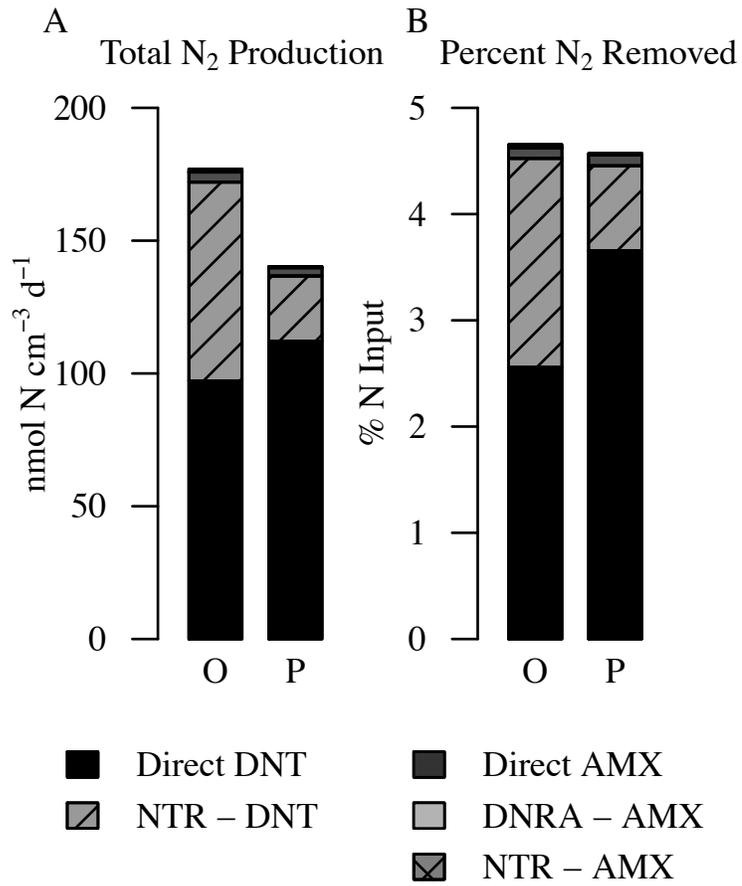

**Figure 7**

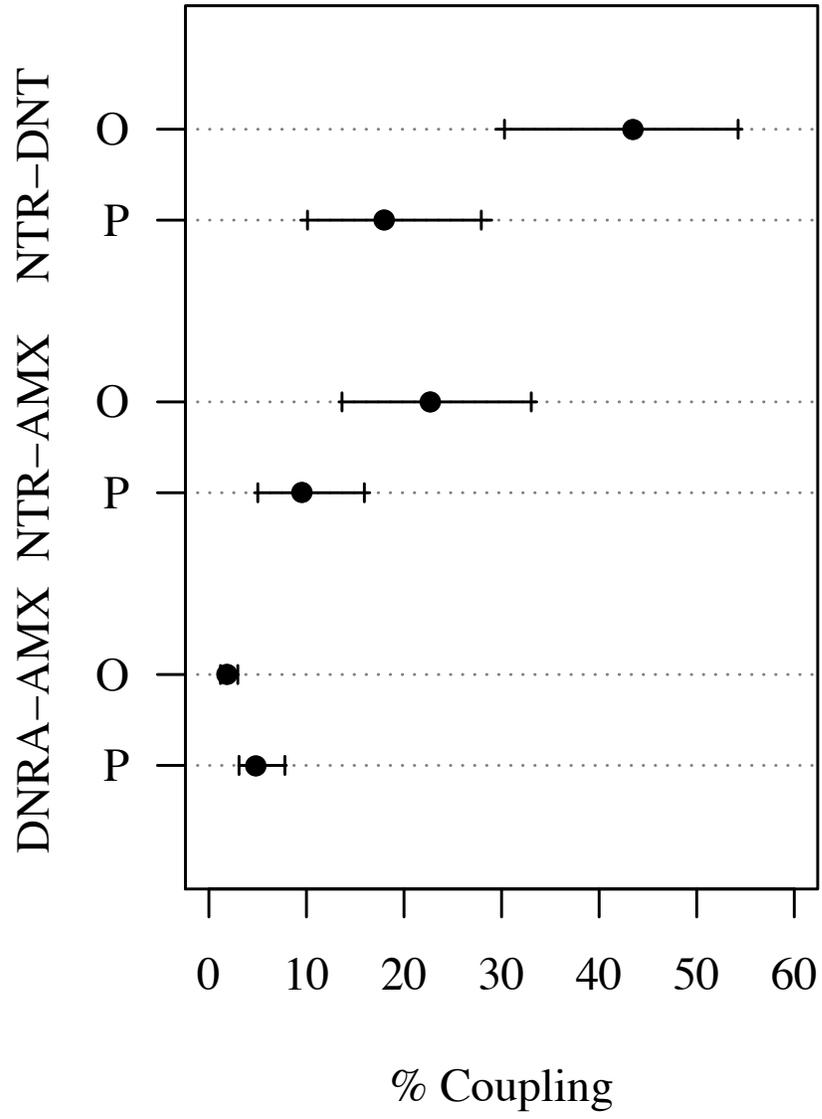

**Figure 8**

**Web Appendix**
*SCOR files*

      Here, we supply model files for two nitrogen cycling networks, one constructed at an oligohaline site and one constructed at a polyhaline site, in the Cape Fear River Estuary, North Carolina. The models are presented in the format specified by the Scientific Committee on Ocean Research standards (SCOR files; Ulanowicz and Kay 1991), and can be utilized by multiple network analysis packages including enaR for R (Lau et al. 2012) and the DOS based software package NETWRK (Ulanowicz and Kay 1991). The models for the oligohaline and polyhaline sites are presented below, as well as in the downloadable files oligo_cfre.dat and poly_cfre.dat files, respectively. The networks for each site contain eight nodes and can be used in most ecological network analyses. The estimation of nitrogen cycling process coupling presented in the manuscript associated with this supplemental information was conducted using expanded fourteen-node models in order to distinguish between multiple outputs from a single node in environ analysis. SCOR files for the fourteen-node versions of each network are available upon request, but were omitted from this publication to avoid confusion. The fourteen-node models artificially inflate the total system throughflow by converting what are boundary fluxes in the eight-node model to internal flows. Subtracting the sum of the boundary flows from the dummy nodes compensates for this inflation and produces identical results to the eight-node model. All network analyses outside of this special case within environ analysis should utilize the eight-node networks for each site.

**Oligohaline SCOR file:**
```
CFRE OLIGOHALINE SUMMER; HINES&LISA&SONG&TOBIAS&BORRETT;
UNITS=NMOLN/CM3/D;
 8  2
W-NH4
W-NOx
W-M
W-ON
S-NH4
S-NOx
S-M
S-ON
 1 1.05e+01
 2 2.76e+01
 3 1.00e-03
 4 5.40e-01
 5 6.83e+02
 6 4.19e+00
 7 1.00e-02
 8 7.11e+03
-1          \IMPORTS
 1 1.30e+02
 2 1.02e+03
```

3 3.90e-05
 4 1.16e+03
 5 1.24e+03
 6 1.73e+02
 7 0.00e+00
 8 7.90e+01
 -1          \EXPORTS
 1 2.76e+02
 2 1.01e+03
 3 3.90e-05
 4 1.16e+03
 5 1.08e+03
 6 1.81e+02
 7 3.90e-10
 8 1.08e+02
 -1          \RESPIRATION
 -1          \FLOWS
 3  1 3.09e+00
 4  1 5.20e+00
 5  1 1.37e+02
 1  2 1.70e+00
 6  2 2.15e+00
 1  3 1.90e+00
 2  3 9.80e+00
 4  3 7.40e+00
 7  3 1.19e+02
 3  4 1.60e+01
 8  4 8.50e+02
 1  5 5.50e+00
 6  5 3.89e+01
 7  5 1.47e+02
 8  5 1.50e+02
 2  6 1.41e+01
 5  6 1.44e+02
 3  7 1.19e+02
 5  7 2.13e+02
 6  7 1.09e+02
 8  7 8.20e+01
 4  8 8.54e+02
 7  8 2.57e+02
-1 -1

**Polyhaline SCOR file:**
CFRE POLYHALINE SUMMER; HINES&LISA&SONG&TOBIAS&BORRETT,
UNITS=NMOLN/CM3/D;
 8  2

W-NH4
W-NOx
W-M
W-ON
S-NH4
S-NOx
S-M
S-ON
 1 2.90e+00
 2 1.51e+01
 3 1.00e-03
 4 4.93e+01
 5 1.08e+02
 6 1.43e+01
 7 1.00e-02
 8 3.72e+03
 -1          \IMPORTS
 1 7.25e+01
 2 3.811e+02
 3 3.90e-05
 4 1.2551e+03
 5 9.751e+02
 6 3.455e+02
 7 0.00e+00
 8 3.91e+01
 -1          \EXPORTS
 1 1.324e+02
 2 3.809e+02
 3 3.90e-05
 4 1.2468e+03
 5 1.0087e+03
 6 2.658e+02
 7 3.90e-07
 8 3.47e+01
 -1          \RESPIRATION
 -1          \FLOWS
 3  1 3.10e+00
 4  1 5.40e+00
 5  1 5.53e+01
 1  2 7.00e-01
 6  2 7.30e+00
 1  3 1.70e+00
 2  3 5.00e-01
 4  3 1.40e+00
 7  3 1.192e+02
 3  4 5.00e-01

```
8  4  4.238e+02
1  5  1.50e+00
6  5  1.044e+02
7  5  1.467e+02
8  5  1.00e+02
2  6  7.70e+00
5  6  7.75e+01
3  7  1.192e+02
5  7  1.862e+02
6  7  5.32e+01
8  7  1.596e+02
4  8  4.258e+02
7  8  2.532e+02
-1 -1
```

**Supplemental Literature Cited**

Lau MK, Borrett SR, Hines DE (2012) enaR: Tools ecological network analysis. R package version 1.01, cran.r-project.org/web/packages/enaR/index.html

Ulanowicz RE, Kay J (1991). A package for the analysis of ecosystem flow networks. Environ Softw 6:131–142.